\documentclass[sigconf]{acmart}
\AtBeginDocument{%
  \providecommand\BibTeX{{%
    \normalfont B\kern-0.5em{\scshape i\kern-0.25em b}\kern-0.8em\TeX}}}

\copyrightyear{2024}
\acmYear{2024}
\setcopyright{rightsretained}
\acmConference[IDE '24]{2024 First IDE Workshop}{April 20, 2024}{Lisbon, Portugal}
\acmBooktitle{2024 First IDE Workshop (IDE '24), April 20, 2024, Lisbon, Portugal}\acmDOI{10.1145/3643796.3648443}
\acmISBN{979-8-4007-0580-9/24/04}

\begin{document}

\title{The Visual Debugger: Past, Present, and Future}

\author{Tim Kr\"{a}uter}
\email{tkra@hvl.no}
\orcid{0000-0003-1795-0611}
\affiliation{
  \institution{Western Norway University of Applied Sciences}
  \city{Bergen}
  \country{Norway}
}

\author{Patrick Stünkel}
\email{patrick.stuenkel@hvl.no}
\orcid{0000-0002-0537-295X}
\affiliation{
  \institution{Western Norway University of Applied Sciences}
  \city{Bergen}
  \country{Norway}
}

\author{Adrian Rutle}
\email{aru@hvl.no}
\orcid{0000-0002-4158-1644}
\affiliation{
  \institution{Western Norway University of Applied Sciences}
  \city{Bergen}
  \country{Norway}
}

\author{Yngve Lamo}
\email{yla@hvl.no}
\orcid{0000-0001-9196-1779}
\affiliation{
  \institution{Western Norway University of Applied Sciences}
  \city{Bergen}
  \country{Norway}
}

\renewcommand{\shortauthors}{Kräuter et al.}
\newcommand{\intellij}{IntelliJ IDEA}

\begin{abstract}
    The Visual Debugger is an IntelliJ IDEA plugin that presents debug information as an object diagram to enhance program understanding.
    Reflecting on our \textit{past} development, we detail the lessons learned and roadblocks we have experienced while implementing and integrating the Visual Debugger into the IntelliJ IDEA.
    Furthermore, we describe recent improvements to the Visual Debugger, greatly enhancing the plugin in the \textit{present}.
    Looking into the \textit{future}, we propose solutions to overcome the roadblocks encountered while developing the plugin and further plans for the Visual Debugger.
\end{abstract}




\received{7 December 2023}
\received[accepted]{25 January 2024}

\maketitle

\renewcommand\UrlFont{\color{blue}}

\section{Introduction}
This paper details the experience of implementing, maintaining, and improving the \textit{Visual Debugger} \cite{krauterVisualDebuggerTool2022}.
The Visual Debugger is available for \intellij{} and Android studio as a plugin \cite{timkrauterVisualDebuggerIntelliJ2024}, however, its architecture makes it adaptable to other Integrated Development Environments (IDEs).
The Visual Debugger automatically hooks into the IDE's debugging process and graphically depicts the current stack frame variables as an \textit{object diagram} to foster program comprehension \cite{krauterVisualDebuggerTool2022}.

In \cite{krauterVisualDebuggerTool2022}, we introduced the Visual Debugger, describing its architecture and a typical usage scenario, as well as comparing it to related tools.
The contributions of this paper are twofold:

\textbf{(1)} We describe key improvements made to the Visual Debugger since \cite{krauterVisualDebuggerTool2022}, see \autoref{sec:visualDebugger}, which aim to enhance program comprehension further and lead to a smoother integration of our plugin into \intellij{}.

\textbf{(2)} We discuss the lessons learned and roadblocks we experienced while developing the Visual Debugger as a plugin for \intellij{}, see \autoref{sec:lessonsLearned}.
In addition, we propose methods to mitigate these roadblocks to achieve smoother and simpler IDE integration in the future from the perspective of plugin and IDE developers. 

\section{The Visual Debugger} \label{sec:visualDebugger}

The Visual Debugger is an open-source IDE plugin that visualizes the stack frame variables as an object diagram to improve program comprehension.
It is available for \intellij{} and Android Studio through the JetBrains Marketplace \cite{timkrauterVisualDebuggerIntelliJ2024, timkrauterVisualDebuggerTool2023} making use of the IntelliJ Platform \cite{kurbatovaIntelliJPlatformFramework2021}.
We integrated our plugin into \intellij{}, the most used Java IDE, with approximately 70\% market share as per the JVM Ecosystem Report 2021 \cite{brianvermeerJVMEcosystemReport2021}.

Desired debugging information might not be present in the top-level variables and has to be obtained by digging multiple levels (following links to related objects) deep into different variables.
Thus, in specific scenarios, especially when data is hierarchically structured, a graphical representation results in a faster and better understanding of the stack frame variables \cite{krauterVisualDebuggerTool2022}.

Until now, we have only received positive feedback regarding the Visual Debugger, which now has close to 8500 downloads\footnote{Last checked on the 22nd of January, 2024, see \cite{timkrauterVisualDebuggerIntelliJ2024}.}.
This marks a more than threefold increase in downloads compared to the initial release of our research paper \cite{krauterVisualDebuggerTool2022} on July 21, 2022, which garnered approximately 2700 downloads.

Additional artifacts, including source code, a demonstration of the Visual Debugger tool, and a description of the Visual Debugging API, can be found in \cite{timkrauterICSE2024Artifacts2024}.

\subsection{Description}

\begin{figure}[ht]
    \centering
    \includegraphics[width=1\linewidth]{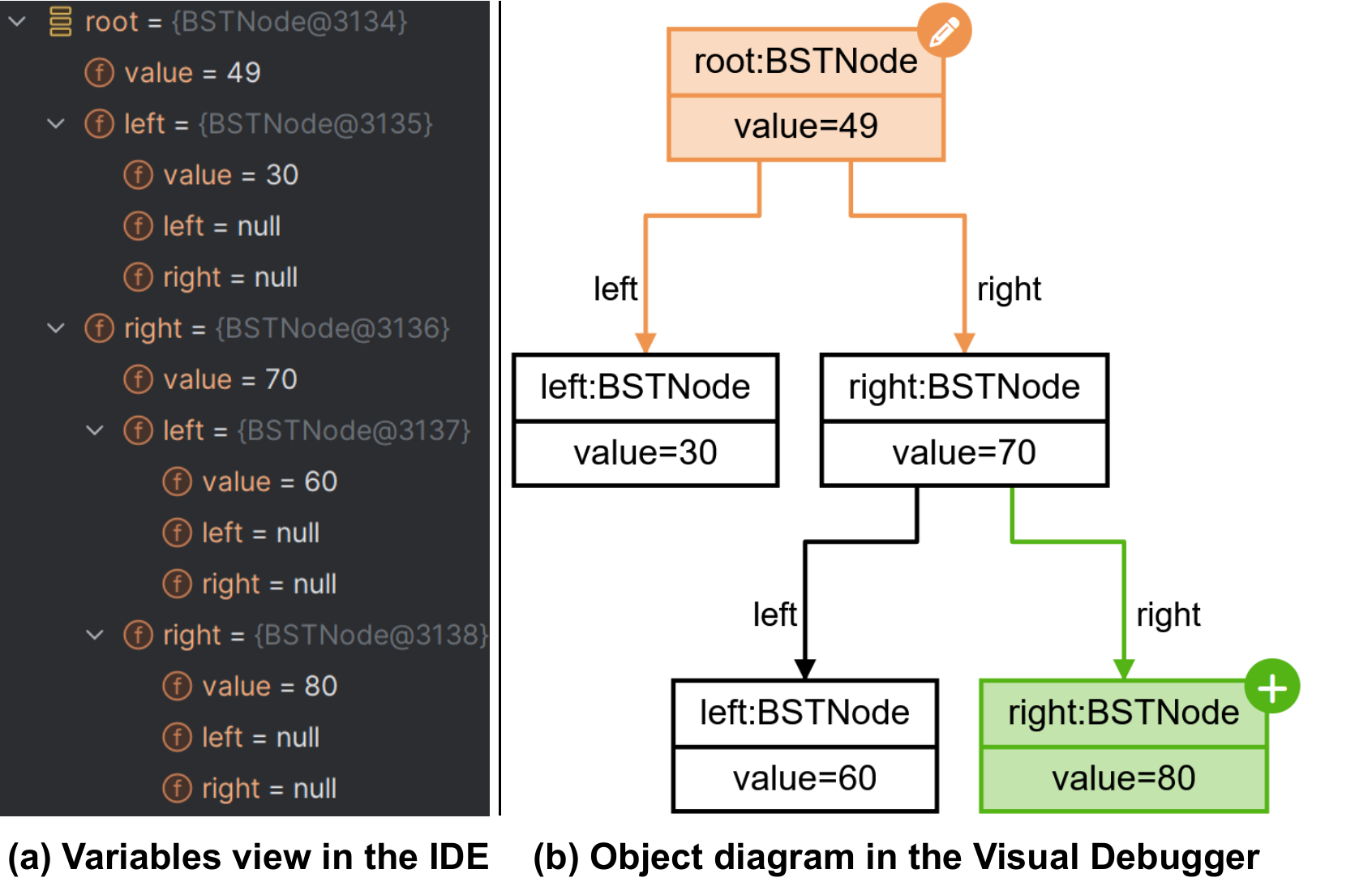}
    \caption{Binary Search Tree (BST) during debugging}
    \label{fig:variables}
\end{figure}

Traditionally, stack frame variables are represented textually, such as in \autoref{fig:variables} (a), a screenshot of the variables view in \intellij{} for a Binary Search Tree (BST).

The Visual Debugger represents the same information graphically as an object diagram.
Using the Visual Debugger is \textit{straightforward} since it opens automatically during debugging in the IDE.
The Visual Debugger visualizes the variables in the scope of the debugging session (see, for example, \autoref{fig:variables} (b)) and, stepping through the source code, updates this representation.
\autoref{fig:variables} (a) and (b) contain the same objects and level of detail, despite \textit{null values} being ignored in the Visual Debugger.
However, the coloring in \autoref{fig:variables} highlights changes and additions, a new feature of the Visual Debugger discussed in \autoref{subsec:improvements}.

The graphical visualization does not replace the textual debugging view but aims to augment the debugging experience to improve program comprehension~\cite{krauterVisualDebuggerTool2022}.
Concretely, the Visual Debugger is \textit{non-intrusive} since it can be used alongside the traditional textual debugging available in the IDE.

The Visual Debugger automatically updates the debug information as the \intellij{} debugger whenever a user steps through the source code or reaches a new breakpoint.
Moreover, the objects a debugging variable refers to can be loaded by double-clicking an object in the object diagram, similar to how it works in most textual debuggers.
For example, in \autoref{fig:variables}, all objects the green object refers to were loaded.
The goal is to make the Visual Debugger \textit{familiar} by adopting how textual debuggers work such that a transition is smooth.

A new video demonstration of the Visual Debugger is available at \url{https://www.youtube.com/watch?v=LsAMTnLxWJw}, showcasing the improvements discussed in \autoref{subsec:improvements}, which were made since our last publication \cite{krauterVisualDebuggerTool2022}.

\subsection{Architecture}
\begin{figure}[ht]
  \centering
  \includegraphics[width=0.7\linewidth]{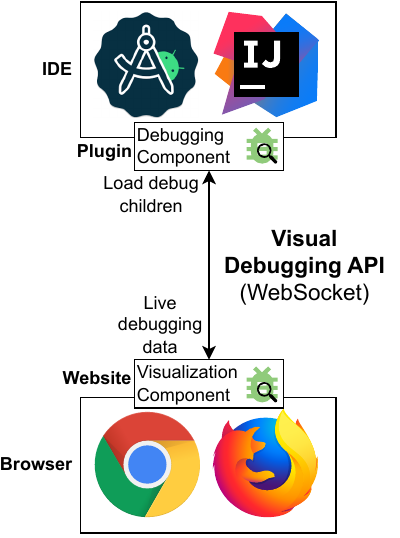}
  \caption{Visual Debugger architecture}
  \label{fig:architecture}
\end{figure}

In this section, we briefly summarize the architecture of the Visual Debugger.
The plugin's architecture plays an important role in understanding the improvements to the Visual Debugger and the roadblocks we discuss in \autoref{sec:lessonsLearned}.
The Visual Debugger is separated into two independent components communicating through the \textit{Visual Debugging API} \cite{krauterVisualDebuggerTool2022}.
\autoref{fig:architecture} summarizes the architecture of the Visual Debugger.

The first component, the \textit{debugging component}, integrates with \intellij{}, and its primary function is to acquire stack frame variables throughout the debugging process repeatably.
The debugging component makes this information available via a WebSocket server that implements our Visual Debugging API.
Upon establishing a connection to the Visual Debugging API, a client receives real-time updates with the latest debugging information and can request to load referenced objects for an existing object.

The second component, the \textit{visualization component}, portrays stack frame variables as an object diagram for better understanding (refer to \autoref{fig:variables} for an illustration).
It is implemented using web technologies and our object diagram library \cite{timkrauterObjectdiagramjs2023} to visualize the debug information.
Additionally, it leverages the Visual Debugging API to communicate with the debugging component.
Thus, it is agnostic of the IDE used for debugging and can even be reused for other programming languages than Java/Kotlin.
In practice, the visualization component is hosted on a web server inside the IDE, as part of the Visual Debugger plugin.

The downside to this flexibility is that the Visual Debugger is not entirely integrated into \intellij{}, as the visualization occurs in a browser external to the IDE.
The rationale behind opting for this approach instead of native IDE integration is discussed in \autoref{sec:lessonsLearned}.
In addition, the next section discusses recent improvements to the Visual Debugger, which led to a better IDE integration despite a web-based user interface.

\subsection{Improvements} \label{subsec:improvements}
We made four major improvements to the Visual Debugger, described in detail in the following section.

\textbf{(1) Browser integration:} We integrated an embedded browser into the Visual Debugger panel inside \intellij{}.
The embedded browser uses the Java Chromium Embedded Framework (JCEF), available by default in \intellij{}.
As a result, users now have the option to use the embedded instead of an external browser.
Most of the functionality of our visualization component worked out of the box with the JCEF browser.
Additionally, nearly all other features, for example, exporting object diagrams to XML/SVG, were successfully implemented using JCEF APIs.

The Chromium Embedded Framework (CEF) \cite{marshallgreenblattChromiumEmbeddedFramework2023} is integrated with numerous programming languages, including the Java integration JCEF.
The CEF plays a crucial role in many popular applications, for example, the cross-platform Steam Client\footnote{\url{https://developer.valvesoftware.com/wiki/Chromium_Embedded_Framework}}.
However, integrating web applications into \intellij{} is not entirely seamless, which is also highlighted by the fact that using JCEF is still an experimental feature in \intellij{}.
Consequently, we give each user the choice to use the embedded or external browser.
We discuss the problems we faced using JCEF and possible improvements in \autoref{sec:lessonsLearned}.

\textbf{(2) Dynamic loading:} We enhanced the loading mechanism for stack frame variables inside the Visual Debugger.
Previously, we had to pre-load and cache more debug information than was requested by the user since we could not load debug information on demand.
The initial load invalidated the underlying stack frame supplied by the Java Debugging Interface (JDI).
Now, we leverage the APIs provided by \intellij{}, which are a thin wrapper around the JDI.
This enables us to defer loading additional debug information until explicitly required.
Consequently, there is no longer a need to pre-load potentially unnecessary debug information.
This optimization has improved performance, making the Visual Debugger applicable to more intricate debugging scenarios.

\textbf{(3) Change highlighting:} The visualization component of the Visual Debugger now \textit{highlights changes} using colors and overlays in the object diagram.
New objects and links are colored green, while changes to existing elements lead to orange coloring.
Computing and highlighting the changes is enabled by default but can be switched off.
\autoref{fig:variables} shows the new change visualization by highlighting one changed and added object accordingly.
A software engineer is usually most interested in the changes that occur to the objects during debugging.
Our color-based visualization of changes in the object diagram makes it easier for a software engineer to see changes even when dealing with a complex debugging situation with multiple connected objects.
Thus, the change highlighting contributes to our goal of enhancing program comprehension.

Changes can be calculated efficiently using unique object IDs provided during debugging.
We implemented the change detection in the visualization component such that it can be reused across programming languages and IDEs \cite{timkrauterICSE2024Artifacts2024}.

\textbf{(4) Debug history:} Furthermore, the visualization component now keeps a \textit{debug history} so that a user can see debug information from previous debugging steps.
The debug history's length is configurable or can be turned off entirely.
As described earlier, software engineers are most interested in how variables change during debugging.
Consequently, to not only highlight differences to the previous step, we save the previous debugging information in a debug history.
A software engineer can thus inspect the previously shown debug information and step as far back as he configured.
The Visual Debugger always shows where the debug information was collected in the source code, and highlights changes compared to the previous debugging step.
In addition, one can still export the debug information to an image or even edit it in our object diagram modeler \cite{timkrauterObjectdiagramjs2023} for documentation purposes \cite{krauterVisualDebuggerTool2022}.

We implemented the \textit{debug history} in the visualization component to be independent of the used programming language and IDE.
Both the \textit{change highlighting} and \textit{debug history} improvements align with our goal to offer a reusable visualization of the stack frame variables~\cite{krauterVisualDebuggerTool2022}.

\section{Lessons Learned \& Roadblocks} \label{sec:lessonsLearned}

A key takeaway from the experiences we gained from developing our plugin is that several useful features and APIs are not or only briefly documented.
Consequently, after conducting an extensive review of the documentation, it proves advantageous to seek assistance by posing questions in the community forum dedicated to plugin development. 
One receives quick and useful responses for clear, detailed inquiries, making the forum invaluable.
For example, our new dynamic loading of debug information discussed in \autoref{subsec:improvements} was enabled due to the \intellij{} forum.
The forum provides a unique opportunity to interact directly with IDE developers.

While developing the Visual Debugger plugin, we encountered two significant roadblocks.
We will now describe these roadblocks and possibilities to overcome them in the future.

\textbf{(1) No native support for web-based UIs:} Creating a user interface (UI) with interactive diagramming capabilities and integrating it into \intellij{} was challenging for different reasons.
The IntelliJ Platform is built on Java, and plugin UIs utilize the Swing framework.
However, we could not find a suitable diagramming framework for Swing to implement an interactive object diagram for our visualization component.
\intellij{} uses \textit{yFiles} \cite{yworksYFilesDiagrammingLibrary2023}, for example, to visualize generated class diagrams from source code.
For us, \textit{yFiles} was not an option because the cheapest license already carries a significant cost, while we do not plan to monetize our plugin.

We started with an embedded visualizer using PlantUML \cite{arnaudroquesPlantUML2023} to generate pictures of object diagrams.
Moreover, to allow interactions with object diagrams, we opted for the free and open-source \textit{diagram-js} library to implement our web-based library \textit{object-diagram-js} \cite{timkrauterObjectdiagramjs2023} used in the Visual Debugger.
To summarize, the rich web ecosystem and its growing popularity among developers have led us to a web-based visualization.
Integrating a web-based UI into \intellij{} is not obvious compared to other IDEs, such as Visual Studio Code, which is built on web technologies and heavily uses web views for extensions.

To integrate our web-based UI into \intellij{}, we now use JCEF as described in \autoref{subsec:improvements}.
We were unaware of this possibility for an extended period.
We believe the plugin documentation regarding JCEF and web views has room for improvement to take full advantage of the rich web ecosystem.
First, it should be more visible that the integration of web views is possible.
Second, the integration poses challenges that can be reduced by more extensive documentation about JCEF.
We authored a pull request to improve the plugin documentation, incorporating insights from our experience with the Visual Debugger.

\textbf{(2) Missing debugging APIs:} Other IDEs than \intellij{} are missing a debugging API to implement our plugin.
Users of the Visual Debugger have asked us if the plugin could be adapted to work with other IDEs and programming languages.
For example, we would like to support debugging \textit{Go} applications inside the GoLand IDE, as requested by a user.
However, compared to \intellij{} plugins, there is no API to hook into debugging processes in GoLand.
Thus, it is currently impossible to adapt the Visual Debugger for GoLand to integrate with the IDE seamlessly.
Our visualization component cannot operate without a debugging component that provides debugging information from GoLand.
This represents a major roadblock to adopting the Visual Debugger for other IDEs and programming languages.
One could develop the necessary APIs and debugging components for each desired IDE and programming language combination.
Nevertheless, there would be significant redundancy across all these implementations.

\begin{figure}[ht]
  \centering
  \includegraphics[width=1\linewidth]{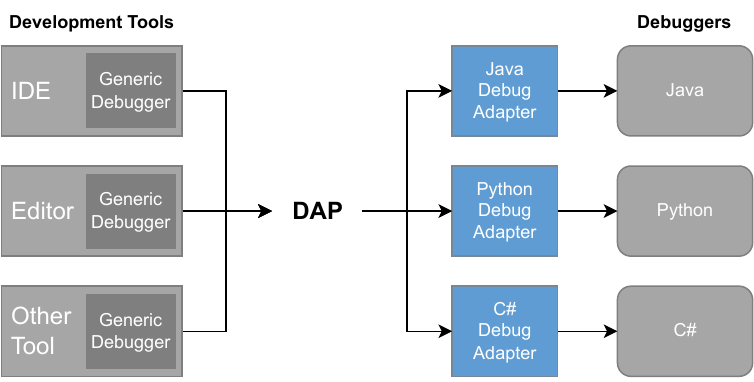}
  \caption{Debug adapter protocol architecture \cite{microsoftDebugAdapterProtocol2023}}
  \label{fig:DAP_Architecture}
\end{figure}

A possible solution would be to utilize the \textit{Debug Adapter Protocol} (DAP) \cite{microsoftDebugAdapterProtocol2023}.
The DAP is a sibling of the more popular \textit{Language Server Protocol} (LSP) \cite{microsoftLanguageServerProtocol2023} and standardizes an abstract protocol for how a development tool communicates with concrete debuggers.
The motivation is that debuggers have to be implemented only once for each language and then can be reused in different IDEs, Editors, or other tools, such as our Visual Debugger, see \autoref{fig:DAP_Architecture}.
Since not all current debuggers will adopt this protocol, an intermediary component is envisioned to adapt an existing debugger to the DAP \cite{microsoftDebugAdapterProtocol2023}, see \textit{debug adapters} in \autoref{fig:DAP_Architecture}.

After the success of the LSP, the DAP could become the next standardized development tool functionality \cite{raskVisualStudioCode2020,borkLanguageServerProtocol2023}.
The official page of the DAP lists 11 tools supporting the DAP, 67 debug adapters, and 11 DAP SDKs as of December 2023 \cite{microsoftDebugAdapterProtocol2023}, including, for example, a Go DAP implementation maintained by Google \cite{googleGoImplementationDebug2023}.
Research is also conducted on the DAP to debug Domain-Specific Languages~\cite{jeanjeanIDECodeReifying2021,enetProtocolBasedInteractiveDebugging2023}.

Like our visual debugging API, the DAP is independent of development tools (as shown on the left side of \autoref{fig:DAP_Architecture}) and is also language-agnostic (as depicted on the right side of \autoref{fig:DAP_Architecture}).
How JetBrains or other IDEs communicate with its integrated debuggers must not be changed to the DAP.
However, each IDE could provide debug adapters for the supported programming languages or integrate existing debug adapters into the IDE.
In addition, using a standard protocol requires minimal documentation and leads to fewer support inquiries.

When the Visual Debugger also supports the DAP in the future, it can attach to the different debug adapters, which all provide the DAP, i.e., the same interface.
As a result, the Visual Debugger becomes compatible with any combination of IDE and programming language that provides a debug adapter.
For example, the Visual Debugger could be automatically available for more IDEs/Editors, such as GoLand (Go), RustRover (Rust), Netbeans (Java), and Visual Studio Code.

\begin{figure}[ht]
  \centering
  \includegraphics[width=1\linewidth]{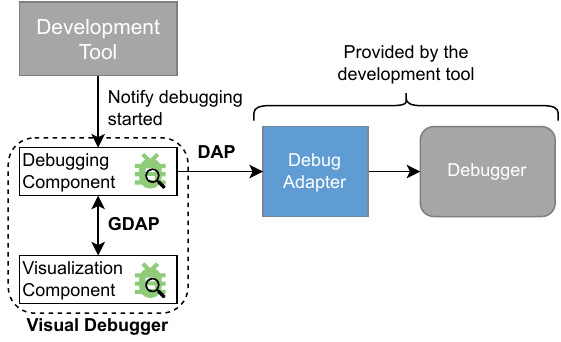}
  \caption{Visual Debugger architecture using the DAP}
  \label{fig:Target_Architecture}
\end{figure}

\autoref{fig:Target_Architecture} shows a possible architecture for the Visual Debugger using the DAP.
Only a light integration for each development tool (IDE, Editor) is needed to get notified about the start of debugging.
Then, the Visual Debugger can use the provided DAP for all other interactions.
The Visual Debugging API can then be thought of as a graphical DAP (GDAP) similar to the graphical LSP (GLSP), an adoption of the LSP for graphical editors \cite{rodriguez-echeverriaLanguageServerProtocol2018,borkVisionFlexibleGLSPBased2024}.

\section{State of the art} \label{sec:relatedWork}
Visual debugging has been researched since the 90s \cite{jerdingUsingVisualizationFoster1994, mukherjeaVisualDebuggingIntegrating1994, hansonSimpleExtensibleGraphical1997}, but most of the resulting tools have yet to become mainstream or integrated with modern IDEs.
We will now present three tools that are still relevant today and compare them to the visual debugger to describe how our contribution fits into the state of the art.

The \textbf{Data Display Debugger (DDD)}\footnote{\url{https://www.gnu.org/software/ddd/}} provides a graphical data visualization as a box-and-pointer diagram \cite{zellerDDDFreeGraphical1996}.
DDD is a graphical standalone debugger that relies on command-line debuggers like the GNU debugger.
Using DDD and the GNU debugger, one can debug executable binaries originally written in C, C\texttt{++}, Go, Rust, and more.
However, other command-line debuggers can be used to debug programs written, for example, in Java or Python.
DDD is a powerful tool and, to our knowledge, the first debugger which includes data visualization.

DDD and the visual debugger are similar since box-and-pointer diagrams are equivalent to object diagrams when switching to the object-oriented programming paradigm.
However, our tool is deeply integrated with \intellij{}, while DDD is a standalone tool used for compiled executable binaries.
Furthermore, the visual debugger incorporates advanced features such as \textit{dynamic loading}, \textit{change highlighting} and a \textit{debug history} which helps deal with complicated debugging scenarios.

\textbf{Java Interactive Visualization Environment (JIVE)}\footnote{\url{https://cse.buffalo.edu/jive/}} is a plugin for the Eclipse IDE \cite{gestwickiJIVEJavaInteractive2004}.
It provides different visualizations during program execution.
Stack frame variables are visualized as an object diagram, while the call history is represented as a compacted UML sequence diagram \cite{jayaramanCompactVisualizationJava2017}.
Furthermore, it allows reverse stepping and even contains features for extracting a finite-state model to allow property checking \cite{k.p.FiniteStateModel2021}.
JIVE provides a custom debugging environment inside the Eclipse IDE to provide its powerful debugging features.

Compared to JIVE, the visual debugger works alongside debugging in the IDE and focuses on object diagram visualization of stack frame variables.
Similar to reverse stepping, we provide the debug history feature.
The visual debugger focuses on simplicity, usability, and seamless integration with the IDE, while JIVE focuses on advanced debugging features that might be harder to understand.
Furthermore, we decouple debugging and visualization to reuse the visualization across different IDEs while JIVE is tied to Java and Eclipse.

The \textbf{Java Visualizer} has been developed as a plugin for the \intellij{} for teaching purposes \cite{lipsitzJavaVisualizerIntelliJ2024}.
Like the visual debugger, it hooks into the debugging process and visualizes the stack frame variables.
Nevertheless, it visualizes variables as a box-and-pointer diagram and depicts all stack frames and their order on the call stack.
The call stack and box-and-pointer diagram are then updated for each debugging step in the IDE.

The Java Visualizer is a helpful tool that has inspired the development of the visual debugger.
However, even in simple situations, the visualization can get complex since it includes all variables from the Java heap, not just the variables within the current stack frame, i.e., the debugging scope.
While this can be an advantage, it often quickly clutters the visualization, especially if one is only interested in the variables within the current scope.
On the contrary, the visual debugger only shows relevant information from the present scope, even filtering the initial depth of the visualization.
In addition, we provide advanced features such as \textit{change highlighting}, which become particularly useful as diagrams grow in size.

\section{Conclusion \& Future work} \label{sec:conclusion}
The Visual Debugger has increased in popularity, measured by the more than tripling of downloads of the plugin compared to our previous publication \cite{krauterVisualDebuggerTool2022}.
In this work, we make two contributions related to the Visual Debugger and the broader topic of IDE integration.

Our first contribution is the new improvements we incorporated into the Visual Debugger.
We integrate our Visual Debugger more smoothly into the IDE by employing an embedded browser based on the Java Chromium Embedded Framework (JCEF) and improve the performance of loading debugging information.
Furthermore, the Visual Debugger now highlights changes graphically using colors and provides a debug history.

Our second contribution is detailing the experience gained by implementing the Visual Debugger.
We describe two major roadblocks hampering tighter IDE integration of our plugin.
First, integrating web-based user interfaces that use the extensive and popular web ecosystem is not trivial.
Using the JCEF makes this integration possible, but its visibility and documentation should be improved.
Second, not all IDEs offer debugging-related APIs such that we could integrate our plugin.
To make debugging functionality uniformly accessible for plugin development, we propose to utilize the standardized Debug Adapter Protocol (DAP).

In future work, we aim to adapt the Visual Debugger for other IDEs and code editors such as GoLand, Eclipse \cite{desrivieresEclipsePlatformIntegrating2004}, and Visual Studio Code since our users have requested this.
We aim to implement the DAP for the Visual Debugger to minimize the integration effort and development cost for the different IDEs.

Furthermore, we aim to study the usability and scalability of the Visual Debugger and its impact on program comprehension.
It seems interesting to find out in which scenarios visual debugging is more effective than textual debugging and vice versa.
Since the Visual Debugger is \textit{not} meant to replace the textual debugger, one should also include a combination of both textual and visual debugging in an empirical study.


\renewcommand\UrlFont{\color{black}}

\bibliographystyle{ACM-Reference-Format}
\bibliography{bib}
\end{document}